\documentclass[pra,showpacs,twocolumn,amsmath,amssymb]{revtex4}
\usepackage{graphicx,bm}
\begin{document}
\title{Entanglement purification of multi-mode quantum states}
\date{\today}
\author{J. Clausen}
\email{J.Clausen@tpi.uni-jena.de}
\author{L. Kn\"oll}
\author{D.-G. Welsch}
\affiliation{Friedrich-Schiller-Universit\"at Jena,\\
             Theoretisch-Physikalisches Institut,\\
             Max-Wien-Platz 1, D-07743 Jena, Germany}
\begin{abstract}
An iterative random procedure is considered allowing an entanglement
purification of a class of multi-mode quantum states. In certain cases, a
complete purification may be achieved using only a single signal state
preparation. A physical implementation based on beam splitter arrays and
non-linear elements is suggested. The influence of loss is analyzed in the
example of a purification of entangled N-mode coherent states.
\end{abstract}
\maketitle
\section{\label{sec1}
       Introduction}
The impossibility to isolate a physical system completely from its environment
represents a major difficulty in the experimental realization of devices
processing quantum information. Of crucial interest are therefore
schemes which tolerate a certain amount of environmental coupling yet protect
the actual information, e.g., by accepting some redundance in the quantum state
preparations. For example, a pure entangled state may be recovered from a number
of mixed entangled states of spatially extended signals shared by separated
parties \cite{entangledBennett}. A potential application of such entanglement
purification schemes is the realization of a secure data transmission despite
using a lossy channel and apparatus \cite{privateBriegel}. Apart from this, the
treatment of entanglement purification helps to deepen the understanding of
entanglement itself \cite{entangledBennett,contPlenio}. Related questions which
have been addressed are whether it is sufficient to process each signal state
copy individually \cite{collectivePopescu} or how to deal with copies which
cannot be factorized into separate states \cite{unknownSchack}. Entanglement
swapping \cite{repeaterZoller,optimalGuo} has been discussed to propose schemes
operating more efficient, despite loss.

While for certain classes of states such a superpositions of coherent states,
methods solely based on linear optical elements like beam splitters and
photodetections could be found \cite{puri1Jeong}, an implementation covering
other classes of entangled states remains a challenge. Schemes based on
cross-Kerr couplers have been described for Gaussian continuous variable
entangled states \cite{gaussian2Zoller,gaussian1Zoller,continuousZoller}.
Alternatively to the generalization to continuous variable states shared by
two stations, multi-user schemes have been considered which involve more than
two parties \cite{multiPlenio}.

The aim of this work is to investigate a range of truncated mixtures of
entangled multi-mode states, whose periodic nature allows a relatively simple
separation of their pure components despite the fact that the states themselves
are infinite. Here, focus is on the theoretical possibility of a purification.
From a practical point of view, an implementation of the
suggested scheme may be realistic only for special signal states. The effect of
loss is discussed in a selected example of application.
\section{\label{sec2}
       Input states}
Within this work, we assume that the mixture describing the input signal state
$\hat{\varrho}$ can be truncated at a sufficiently large number $M$,
\begin{equation}
  \hat{\varrho}=\sum_{n=0}^Mp_n|\Psi_n\rangle\langle\Psi_n|,
  \qquad
\langle\Psi_{n^\prime}|\Psi_n\rangle=\delta_{n^\prime n}.
\end{equation}
By entanglement purification we understand the extraction of some pure
component $|\Psi_0\rangle\langle\Psi_0|$ of $\hat{\varrho}$ by effectively
implementing a transformation
$\hat{\varrho}^\prime$ $\!=$ $\!p^{-1}\hat{Y}\hat{\varrho}\hat{Y}^\dagger$,
$p$ $\!=$ $\!\bigl\langle\hat{Y}^\dagger\hat{Y}\bigr\rangle$ with
$\hat{Y}$ $\!\sim$ $\!|\Psi_0\rangle\langle\Psi_0|$. It may therefore be
understood as a multi-mode state purification. We assume that in each signal
mode $j$, whose entity is denoted by $\bm{I}$ $\!=$ $\!1,\ldots,N$, a
repeater is located whose operation involves some auxiliary modes $\bm{b}_j$. If
the repeaters can only use local preparations $|F\rangle_{\bm{b}_j}$, unitary
transformations $\hat{U}_{j,\bm{b}_j}$, and detections
$_{\bm{b}_j}\langle G_j|$, they can solely implement local operators
$\hat{Y}_j(G_j)$ $\!=$ $\!\hat{U}_j(G_j)
\;_{\bm{b}_j}\langle G_j|\hat{U}_{j,\bm{b}_j}|F\rangle_{\bm{b}_j}$,
which may depend on the results $G_j$ of the local detections in the auxiliary
modes. In contrast, the $|\Psi_n\rangle$ may be arbitrary entangled states of
the spatially separated signal modes. Therefore, the repeaters must share an
entangled state in general. If this state has not been provided previously by a
separate resource similar to teleportation setups, then the only alternative is
to equip each of the repeaters with a second signal input mode $-j$, whose
entity is denoted by $\bm{II}$ $\!=$ $\!-1,\ldots,-N$, and which is fed with a
second copy of the signal state $\hat{\varrho}$. If the repeaters periodically
receive preparations $\hat{\varrho}$, the purification can be implemented in a
loop (using, e.g., a cavity) as a repeated transformation of the signal, each
time applying a further copy of the signal state.

In what follows, we assume that there exists a single-mode basis
$\{|\Phi_n\rangle\}$ allowing the representation of the signal eigenstates in
the form of
\begin{eqnarray}
  |\Psi_{n}\rangle&=&(M+1)^{\frac{1-N}{2}}\sum_{\bm{l}=0}^\infty
  c_{\bm{l}}(n)\sum_{\bm{n}}\,^{(n)}
  \nonumber\\
\label{eigenstates}
  &&\times\;|\Phi_{l_1(M+1)+n_1}\rangle_1\cdots|\Phi_{l_N(M+1)+n_N}\rangle_N,
\end{eqnarray}
where $\bm{l}$ $\!=$ $\!l_1,\ldots,l_N$, and to ensure normalization we must
have $\sum_{\bm{l}=0}^\infty|c_{\bm{l}}(n)|^2$ $\!=$ $\!1$.
The expression $\sum_{\bm{n}}^{(n)}$ denotes the sum over all
$(M$ $\!+$ $\!1)^{N-1}$ different $N$-tuples
$\bm{n}$ $\!=$ $\!n_1,\ldots,n_N$ of numbers
$n_j$ that can each take the values $0,\ldots,M$ but must obey
$\bigl\lceil\sum_{j=1}^Nn_j\bigr\rceil$ $\!=$ $\!n$. Here, we have used the
notation
\begin{equation}
  \lceil m\rceil \equiv \mathrm{Mod}(m,M+1).
\end{equation}
In order to explain the operation of the purification scheme, it will be
convenient to distinguish between the states
\begin{subequations}
\label{varrho-a}
\begin{eqnarray}
\label{varrhoa}
  \hat{\mathcal{R}}_{\bm{I}}&=&\sum_{n=0}^MP_n
  |\Psi_n\rangle_{\bm{I}}\langle\Psi_n|,
  \\
\label{varrhob}
  \hat{\varrho}_{\bm{II}}&=&\sum_{n=0}^Mp_n
  |\Psi_n\rangle_{\bm{II}}\langle\Psi_n|,
\end{eqnarray}
\end{subequations}
where $\hat{\mathcal{R}}_{\bm{I}}$ is the state of the signal currently
processed in the loop and $\hat{\varrho}_{\bm{II}}$ that of the original signal
just entering the second signal input ports of the repeaters.
\section{\label{sec3}
       Implementation of the quantum repeater}
\begin{figure}[ht]
\includegraphics[width=5.1cm]{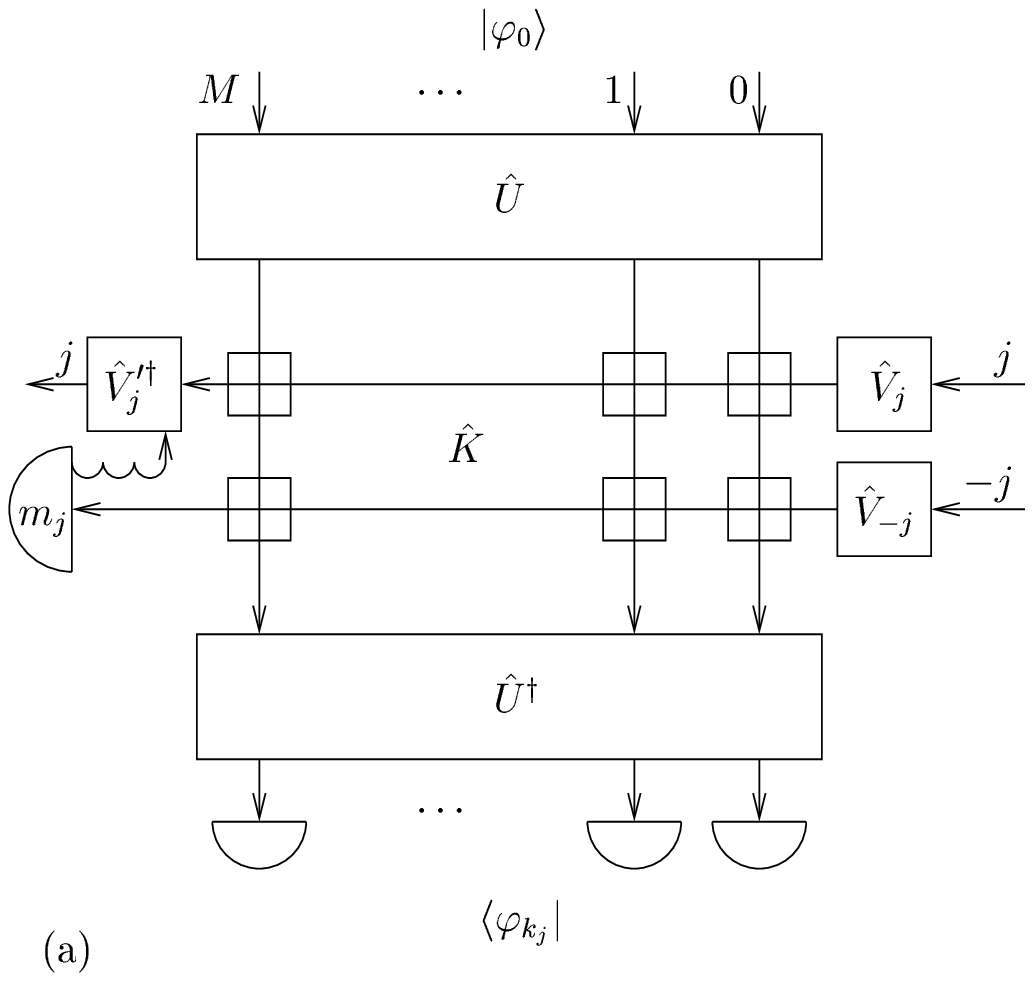}
\hfill\includegraphics[width=2.9cm]{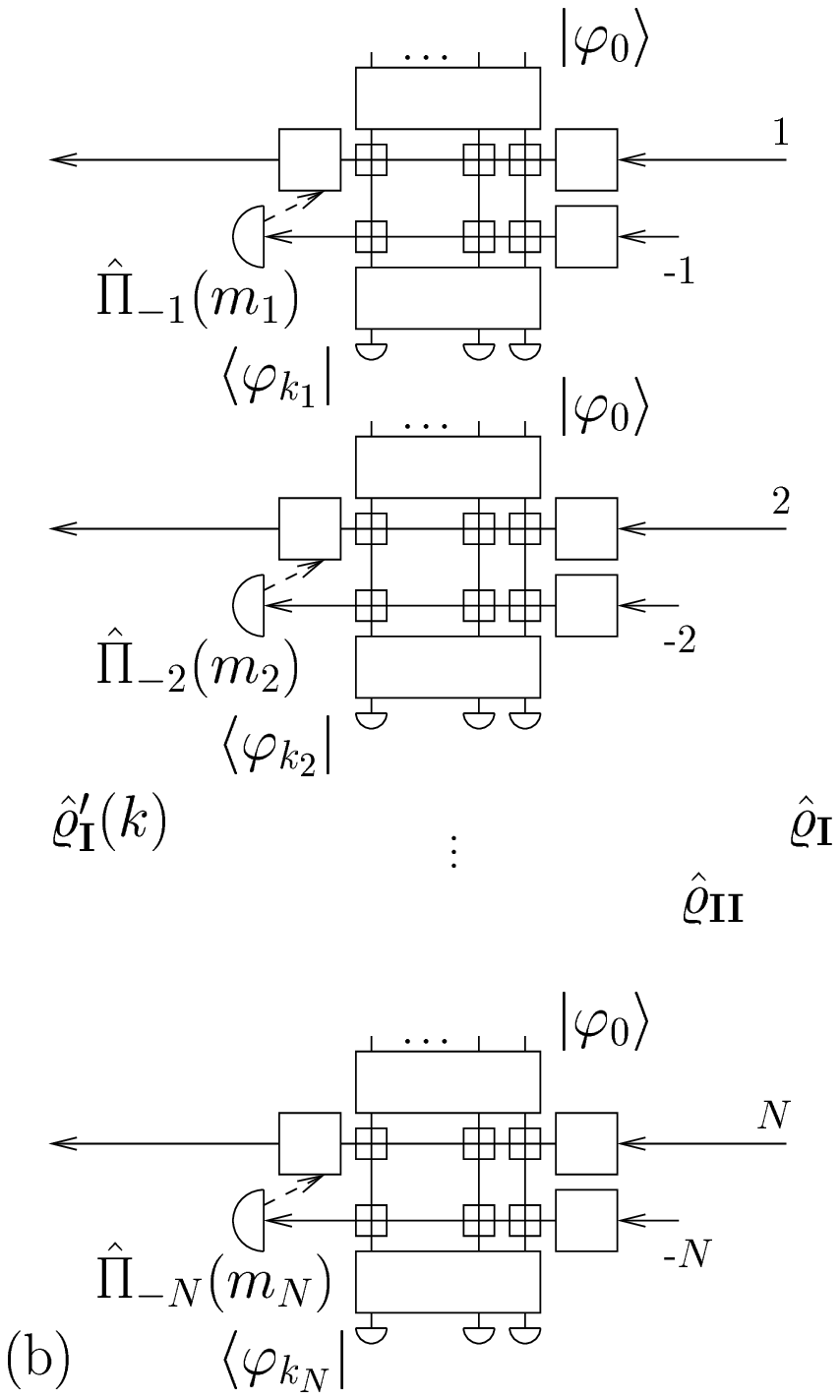}
\caption{\label{fig1}
  Implementation of an $N$-mode entanglement purification. The overall setup
  shown in (b) consists of quantum repeaters located in each of the $N$ signal
  modes $\bm{I}$ $\!=$ $\!1,\ldots,N$. Their operation requires a second copy
  of the signal state to be fed into the inputs
  $\bm{II}$ $\!=$ $\!-1,\ldots,-N$. A detailed view of a single repeater is
  given in (a).
}
\end{figure}

To describe the transformation of the input states Eqs.~(\ref{varrho-a}), let us
take a closer look at the quantum repeater illustrated in Fig.~\ref{fig1}(a).
$\hat{V}$ is a unitary single-mode operator that transforms the basis states
$|\Phi_n\rangle$ to Fock states $|n\rangle$ $\!=$ $\!\hat{V}|\Phi_n\rangle$,
i.e., the single-mode basis $\{|\Phi_n\rangle\}$ is supposed to be known. A
realization of a desired single-mode operator or detection in a truncated space
also based on beam splitter arrays, zero and single photon detections as well as
cross-Kerr elements is presented in \cite{clausen9}. (Note that the $\hat{V}$
are not required if attention is limited to the photon number basis.) Enclosed
between devices implementing $\hat{V}$ and $\hat{V}^\dagger$, an array of
2($M$+1) cross-Kerr elements is placed, whose operation is described by the
operator
\begin{equation}
  \hat{K}=\mathrm{e}^{\mathrm{i}\frac{2\pi}{M+1}
  (\hat{a}_{j}^\dagger\hat{a}_{j}^{}\,-\,\hat{a}_{-j}^\dagger\hat{a}_{-j}^{})
  \sum_{m=0}^Mm\hat{b}_m^\dagger\hat{b}_m^{}}.
\end{equation}
This array couples the signal modes $j$ and $-j$ to $M$+1 auxiliary modes
$0,\ldots,M$, themselves coupled by two 2($M$+1)-port beam splitter arrays
implementing the operators $\hat{U}$ and $\hat{U}^\dagger$, respectively. The
latter are defined by the matrix elements
\begin{subequations}
\begin{eqnarray}
  U_{kl}&=&\langle\varphi_k|\hat{U}|\varphi_l\rangle
  =\frac{1}{\sqrt{M+1}}\mathrm{e}^{\mathrm{i}kl\frac{2\pi}{M+1}},
  \\
  |\varphi_k\rangle&=&\hat{b}_k^\dagger|0\rangle_0\cdots|0\rangle_M.
\end{eqnarray}
\end{subequations}
Initially, these auxiliary modes are prepared in the state $|\varphi_0\rangle$,
i.e., a single-excited Fock state is fed into input port 0 of the
$\hat{U}$-array, while the remaining input ports are left in the vacuum.
Assuming a final detection of the auxiliary modes in the state
$|\varphi_{k_j}\rangle$, i.e., the photodetectors at the output ports $m$ of the
$\hat{U}^\dagger$-array detect $\delta_{m,k_j}$ photons, the action of the $j$th
repeater without the detector placed in the signal output port $-j$ can be
formally described by an operator
\begin{subequations}
\begin{eqnarray}
\label{Yj}
  \hat{Y}_{j,-j}(k_j)&=&\hat{V}_{j}^{\prime\dagger}
  \langle\varphi_{k_j}|\hat{U}^\dagger\hat{K}\hat{U}|\varphi_0\rangle
  \hat{V}_{-j}\hat{V}_{j},
  \\
  \langle\varphi_{k_j}|\hat{U}^\dagger\hat{K}\hat{U}|\varphi_0\rangle
  &=&\frac{1}{M+1}\sum_{m=0}^M
  \nonumber\\
  &&\times\;\mathrm{e}^{\mathrm{i}\frac{2\pi}{M+1}
  (\hat{a}_{j}^\dagger\hat{a}_{j}^{}\,-\,\hat{a}_{-j}^\dagger\hat{a}_{-j}^{}
  \,-\,k_j)m}.\;\;\;
\end{eqnarray}
\end{subequations}
The detector placed in the signal output port $-j$ is assumed to perform a
Pegg-Barnett phase measurement such that each trial gives a value $m_j$ that can
take the values $0,\ldots,M$ and is described by the Positive Operator Valued
Measure (POVM)
\begin{subequations}
\begin{eqnarray}
\label{POVMja}
  \hat{\Pi}_{-j}(m_j)&=&\sum_{l=0}^\infty|l,m_j\rangle\langle l,m_j|,
  \\
  |l,m_j\rangle&=&\frac{1}{\sqrt{M+1}}
  \sum_{n=0}^M\mathrm{e}^{\mathrm{i}\frac{2\pi}{M+1}m_jn}
  \nonumber\\
  &&\times\;|l(M+1)+n\rangle_{-j}.
\end{eqnarray}
\end{subequations}
In turn, the value $m_j$ determines the operator
\begin{equation}
  \hat{V}_{j}^{\prime\dagger}
  =\hat{V}_{j}^\dagger
  \mathrm{e}^{\mathrm{i}\frac{2\pi}{M+1}m_j\hat{a}_{j}^\dagger\hat{a}_{j}^{}}
\end{equation}
in Eq.~(\ref{Yj}).
\section{\label{sec4}
       Input-output relations}
We now consider the overall setup including all signal modes as depicted in
Fig.~\ref{fig1}(b). If states $|\varphi_{k_j}\rangle$ and POVM values $m_j$
are detected in the auxiliary and second signal output modes of the quantum
repeaters, the reduced state of the first signal output modes becomes
\begin{eqnarray}
  \hat{\mathcal{R}}^\prime_{\bm{I}}(\bm{m},\bm{k})
  &=&\frac{\mathrm{Tr}_{\bm{II}}\left[\hat{Y}(\bm{k})
  \hat{\mathcal{R}}_{\bm{I}}\otimes\hat{\varrho}_{\bm{II}}
  \hat{Y}^\dagger(\bm{k})\hat{\Pi}(\bm{m})\right]}{p(\bm{m},\bm{k})}
  \nonumber\\
\label{pretrafo}
  &=&\frac{(M+1)^{1-2N}}{p(\bm{m},\bm{k})}\hat{\mathcal{R}}_{\bm{I}}
  \hat{W}^{\dagger k}
  \hat{\varrho}_{\bm{I}}
  \hat{W}^k.
\end{eqnarray}
In the first line of Eq.~(\ref{pretrafo}), $\bm{k}$ $\!=$ $\!k_1,\ldots,k_N$ and
\begin{equation}
  \hat{Y}(\bm{k})=\prod_{j=1}^N\hat{Y}_{j,-j}(k_j)
\end{equation}
denotes the product of the individual operators Eq.~(\ref{Yj}). Similarly,
$\bm{m}$ $\!=$ $\!m_1,\ldots,m_N$ and
\begin{equation}
  \hat{\Pi}(\bm{m})=\prod_{j=1}^N\hat{\Pi}_{-j}(m_j)
\end{equation}
is the product of the respective single-mode POVM's Eq.~(\ref{POVMja}). In the
second line of Eq.~(\ref{pretrafo}), an operator $\hat{W}$ has been introduced
to express a rotation of the eigenstate indices according to
$\hat{W}^\dagger|\Psi_n\rangle$ $\!=$ $\!|\Psi_{\lceil n+1\rceil}\rangle$,
and $k$ $\!=$ $\!\bigl\lceil\sum_{j=1}^Nk_j\bigr\rceil$.
$\hat{\varrho}_{\bm{I}}$ is the original input state $\hat{\varrho}_{\bm{II}}$
given by Eq.~(\ref{varrhob}) but transferred from $\bm{II}$ to $\bm{I}$, i.e.,
using a formal mode transfer operator
$\hat{I}_{\bm{I},\bm{II}}$ $\!=$ $\!\prod_{j=1}^N\hat{I}_{j,-j}$, where
\begin{equation}
  \hat{I}_{jk}=\sum_{n=0}^{\infty}|n\rangle_j\,_k\langle n|
  =\,_k\langle0|\mathrm{e}^{\hat{a}_j^\dagger\hat{a}_k^{}}|0\rangle_j
  =\hat{I}_{kj}^\dagger,
\end{equation}
we may write
$\hat{\varrho}_{\bm{I}}^{}$ $\!=$
$\!\hat{I}_{\bm{I},\bm{II}}^{}\hat{\varrho}_{\bm{II}}^{}
\hat{I}_{\bm{I},\bm{II}}^\dagger$. Since only the signal modes $\bm{I}$ appear
in the second line of Eq.~(\ref{pretrafo}), let us leave out this mode index in
what follows. From Eq.~(\ref{pretrafo}), we see that the state
$\hat{\mathcal{R}}^\prime(\bm{m},\bm{k})$ only depends on $k$, so that the
signal output state in case of an event $k$ can be written as
\begin{eqnarray}
  \hat{\mathcal{R}}^\prime(k)
  &=&\frac{1}{p(k)}\sum_{\bm{m}=0}^M\sum_{\bm{k}}\,^{(k)}\;\;
  p(\bm{m},\bm{k})\hat{\mathcal{R}}^\prime(\bm{m},\bm{k})
  \nonumber\\
\label{trafo}
  &=&\frac{1}{p(k)}\hat{\mathcal{R}}
  \hat{W}^{\dagger k}\hat{\varrho}\hat{W}^k.
\end{eqnarray}
A repeated run of the purification can then be described as an iteration such
that the $j$th iteration step generates a signal state
\begin{equation}
\label{iteration}
  \hat{\mathcal{R}}^{(j)}(k)
  =\frac{1}{p(k)}\hat{\mathcal{R}}^{(j-1)}
  \hat{W}^{\dagger k}\hat{\varrho}\hat{W}^{k}
\end{equation}
from its precursor $\hat{\mathcal{R}}^{(j-1)}$, originally descending from a
signal state copy, $\hat{\mathcal{R}}^{(0)}$ $\!=$ $\!\hat{\varrho}$.
At each step, $k$ $\!=$ $\!k^{(j)}$ takes some value $0,\ldots,M$. An observer
unaware of any detection result simply observes the average
\begin{equation}
\label{average}
  \hat{\mathcal{R}}^{(j)}=\sum_{k=0}^Mp(k)\hat{\mathcal{R}}^{(j)}(k)
  =\hat{\mathcal{R}}^{(j-1)},
\end{equation}
so that $\hat{\mathcal{R}}^{(j)}$ $\!=$ $\!\hat{\mathcal{R}}^{(0)}$ $\!=$
$\!\hat{\varrho}$, which is just the unchanged original signal state. Thinking
of the complete purification procedure as a linear array of repeating stations
rather than a loop, the resulting setup therefore constitutes a
\textquoteleft{quantum state guide}\textquoteright\, which only may increase the
observer's knowledge rather than changing the signal state. In the special case
of an equipartition, $p_n$ $\!=$ $\!(M$ $\!+$ $\!1)^{-1}$, no additional
knowledge can be gained even from the measurements, so that a purification is
not possible, $\hat{\mathcal{R}}^{(j)}(k)$ $\!=$ $\!\hat{\mathcal{R}}^{(j-1)}$.
By taking into account all detection results $k^{(1)},\ldots,k^{(j)}$ obtained
during the steps $1,\ldots,j$, the explicit expression of the iterated state
becomes
\begin{eqnarray}
  \hat{\mathcal{R}}^{(j)}\left[k^{(1)},\ldots,k^{(j)}\right]
  &=&\frac{\hat{\varrho}\prod_{l=1}^j
  \hat{W}^{\dagger k^{(l)}}\hat{\varrho}\hat{W}^{k^{(l)}}}
  {p\left[k^{(1)},\ldots,k^{(j)}\right]}
  \nonumber\\
\label{iteratedstate}
  &=&\sum_{n=0}^MP^{(j)}_n|\Psi_n\rangle\langle\Psi_n|
\end{eqnarray}
with
\begin{eqnarray}
  P^{(j)}_n&=&\frac{1}{p\left[k^{(1)},\ldots,k^{(j)}\right]}
  p_n\prod_{l=1}^jp_{\lceil n-k^{(l)}\rceil}
  \nonumber\\
\label{iteratedp}
  &=&\frac{1}{p\left[k^{(1)},\ldots,k^{(j)}\right]}
  p_n\prod_{l=0}^Mp_{\lceil n-l\rceil}^{s_l},
\end{eqnarray}
where $s_l$ is the number of events for which $k$ $\!=$ $\!l$. From
Eq.~(\ref{iteratedp}) we see that $\hat{\mathcal{R}}^{(j)}$ only depends on
$\bm{s}$ $\!=$ $\!s_0,\ldots,s_M$, so that we can refer the iterated state to
this event,
\begin{eqnarray}
  &&\hat{\mathcal{R}}^{(j)}(\bm{s})
  \nonumber\\
  &=&\frac{1}{p(\bm{s})}\sum_{k^{(1)},\ldots,k^{(j)}}^*
  p\left[k^{(1)},\ldots,k^{(j)}\right]
  \hat{\mathcal{R}}^{(j)}\left[k^{(1)},\ldots,k^{(j)}\right]
  \nonumber\\
  &=&\frac{1}{p(\bm{s})}\binom{j}{\bm{s}}\sum_{n=0}^M
  p_n\prod_{l=0}^Mp_{\lceil n-l\rceil}^{s_l}|\Psi_n\rangle\langle\Psi_n|
  \nonumber\\
\label{finalstate}
  &=&\frac{1}{p(\bm{s})}\binom{j}{\bm{s}}
  \hat{\varrho}^{s_0+1}\hat{W}^\dagger\hat{\varrho}^{s_1}\hat{W}^\dagger
  \hat{\varrho}^{s_2}\hat{W}^\dagger
  \cdots\hat{W}^\dagger\hat{\varrho}^{s_M}\hat{W}^\dagger.
\end{eqnarray}
Here $\sum^*$ denotes the sum over all $j$-tuples $k^{(1)},\ldots,k^{(j)}$
containing $s_l$ times the numbers $l$ ($\!=$ $\!0,\ldots,M$), whose number
is given by the polynomial coefficient $\binom{j}{\bm{s}}$. Since a pure signal
input state results in a pure signal output state, the random walk of the state
may eventually end in a pure state. If this happens, the probability of
obtaining $|\Psi_n\rangle$ is
$p_n$ $\!=$ $\!\langle\Psi_n|\hat{\varrho}|\Psi_n\rangle$ as becomes plausible
from Eq.~(\ref{average}).
\section{\label{sec5}
       Binary operation}
\begin{figure}[ht]
\includegraphics[width=8.6cm]{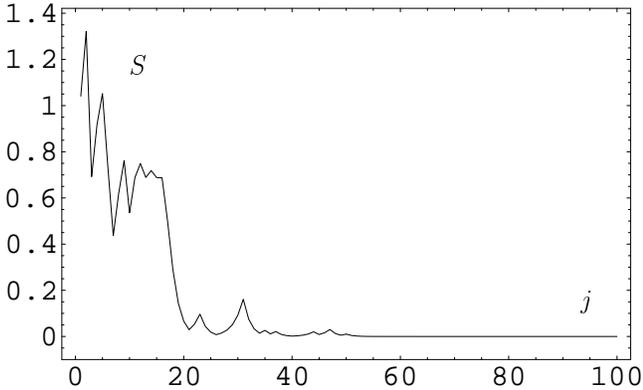}
\caption{\label{fig2}
  Typical evolution of the von Neumann entropy
  $S$ $\!=$ $\!-\mathrm{Tr}[\hat{\mathcal{R}}^{(j)}\ln\hat{\mathcal{R}}^{(j)}]$
  of the signal state $\hat{\mathcal{R}}^{(j)}$ as seen by an observer aware of
  the measurement results occuring during the first $j$ cycles of a left alone
  (random) purification process according to Eqs.~(\ref{biniteration-a}). The
  initial probabilities are assumed to be
  $p_n$ $\!\sim$ $\!\mathrm{e}^{-n}$, where $n$ $\!=$ $\!0,\ldots,100$.
}
\end{figure}

Let us return to Eq.~(\ref{iteration}). Since we are only interested in
obtaining the state $|\Psi_0\rangle$, we may limit attention to a binary
detection by only discriminating between the event $k$ $\!=$ $\!0$ and its
complement $k$ $\!\neq$ $\!0$. Eq.~(\ref{iteration}) then gives the
transformation
\begin{subequations}
\label{biniteration-a}
\begin{eqnarray}
\label{biniterationa}
  \hat{\mathcal{R}}^{(j)}(0)&=&\frac{1}{p(0)}\hat{\mathcal{R}}^{(j-1)}
  \hat{\varrho},
  \\
\label{biniterationb}
  \hat{\mathcal{R}}^{(j)}(\neg0)
  &=&\frac{1}{1-p(0)}\hat{\mathcal{R}}^{(j-1)}(\hat{I}-\hat{\varrho}).
\end{eqnarray}
\end{subequations}
Analogous to Eq.~(\ref{finalstate}), the explicit expression of the iterated
state
\begin{equation}
  \hat{\mathcal{R}}^{(j)}(q)=\frac{1}{p(q)}\binom{j}{q}
  \hat{\varrho}^{q+1}(\hat{I}-\hat{\varrho})^{j-q}
\end{equation}
can be expressed in terms of the number $q$ of events $k$ $\!=$ $\!0$. To give
an example, in a situation where our desired state $|\Psi_0\rangle$ is still
dominant over the impurities,
$p_0$ $\!>$ $\!p_n$ $\!\forall$ $\!n$ $\!\neq$ $\!0$, the state $|\Psi_0\rangle$
is obtained in situations when $j^{-1}q$ is sufficiently large. This case is
illustrated in Fig.~\ref{fig2}, monitoring on the basis of the expectation
value of the von Neumann entropy
$\hat{S}$ $\!=$ $\!-\ln\hat{\mathcal{R}}^{(j)}$ a successful example of a
simulated random purification process calculated according to
Eqs.~(\ref{biniteration-a}).
\section{\label{sec6}
       Purification in a single instant}
If the signal state is a mixture of the $(M+1)^N$ states
\begin{equation}
\label{eigenstates2}
  |\Psi_{\bm{n}}\rangle=\sum_{\bm{l}=0}^\infty
  c_{\bm{l}}(\bm{n})
  |\Phi_{l_1(M+1)+n_1}\rangle_1\cdots|\Phi_{l_N(M+1)+n_N}\rangle_N
\end{equation}
instead of Eq.~(\ref{eigenstates}), each copy can be purified individually
without need of further inputs. All we have to do is feeding the second signal
input ports of the repeaters with vacuum states $|0\rangle_{-j}$ and removing
the devices implementing $\hat{V}_{-j}$ thus making the $m_j$-measurements
redundant. Together with $\hat{\varrho}_{\bm{I}}$ $\!=$
$\!\sum_{\bm{n}=0}^Mp_{\bm{n}}|\Psi_{\bm{n}}\rangle\langle\Psi_{\bm{n}}|$,
Eq.~(\ref{pretrafo}) is then replaced with
\begin{eqnarray}
  \hat{\varrho}^\prime_{\bm{I}}(\bm{k})
  &=&\frac{\mathrm{Tr}_{\bm{II}}\left[\hat{Y}(\bm{k})
  \hat{\varrho}_{\bm{I}}\otimes|\bm{0}\rangle_{\bm{II}}\langle\bm{0}|
  \hat{Y}^\dagger(\bm{k})\right]}{p(\bm{k})}
  \nonumber\\
\label{pretrafo2}
  &=&|\Psi_{\bm{k}}\rangle\langle\Psi_{\bm{k}}|.
\end{eqnarray}
The probability of obtaining $|\Psi_{\bm{k}}\rangle$ is just $p_{\bm{k}}$.
\section{\label{sec7}
       Influence of loss in an example}
Let us return to mixtures of states Eq.~(\ref{eigenstates}). From a practical
point of view, the loss occuring in the purification loop must be taken into
account. The problem is that its effect depends on the respective circulating
signal state $\hat{\mathcal{R}}$ itself. Let us therefore choose the particular
example of a superposition
\begin{eqnarray}
  \hat{\varrho}(r)&=&\frac{\bigl(|\bm{\alpha}\rangle\langle\bm{\alpha}|
  +|\!-\!\bm{\alpha}\rangle\langle\!-\!\bm{\alpha}|\bigr)
  +r\bigl(|\bm{\alpha}\rangle\langle\!-\!\bm{\alpha}|
  +|\!-\!\bm{\alpha}\rangle\langle\bm{\alpha}|\bigr)}
  {2(1+r\mathrm{e}^{-2|\bm{\alpha}|^2})}
  \nonumber\\
\label{applstate}
  &=&\sum_{n=0}^1p_n|\Psi_n\rangle\langle\Psi_n|
\end{eqnarray}
of two $N$-mode coherent states $|\bm{\alpha}\rangle$ $\!=$
$\!|\alpha_1\rangle_1\cdots|\alpha_N\rangle_N$, characterized by their complex
amplitude $\bm{\alpha}$ $\!=$ $\!\alpha_1,\ldots,\alpha_N$ according to
$|\alpha_j\rangle_j$ $\!=$
$\!\mathrm{e}^{\alpha_j\hat{a}_j^\dagger-\alpha_j^*\hat{a}_j}|0\rangle_j$. The
eigenvalues and eigenstates are given by
\begin{subequations}
\begin{eqnarray}
  p_n&=&\frac{1+(-1)^n\mathrm{e}^{-2|\bm{\alpha}|^2}}
  {2(1+r\mathrm{e}^{-2|\bm{\alpha}|^2})}[1+(-1)^nr],
  \\
  |\Psi_n\rangle&=&\frac{|\bm{\alpha}\rangle+(-1)^n|\!-\!\bm{\alpha}\rangle}
  {\sqrt{2[1+(-1)^n\mathrm{e}^{-2|\bm{\alpha}|^2}]}},
\end{eqnarray}
\end{subequations}
where we have defined $|\bm{\alpha}|^2$ $\!=$ $\!\sum_{j=1}^N|\alpha_j|^2$. In
the mesoscopic case, $\mathrm{e}^{-2|\alpha_j|^2}$ $\!\ll$ $\!1$, the
$|\Psi_n\rangle$ can be written in the form of Eq.~(\ref{eigenstates}),
\begin{subequations}
\begin{eqnarray}
  |\Psi_n\rangle
  &\approx&2^{\frac{1-N}{2}}
  \sum_{\bm{n}}\,^{(n)}\;\;|\Phi_{n_1}\rangle\cdots|\Phi_{n_N}\rangle,
  \\
  |\Phi_{n_j}\rangle
  &=&\frac{|\alpha_j\rangle+(-1)^{n_j}|\!-\!\alpha_j\rangle}
  {\sqrt{2[1+(-1)^{n_j}\mathrm{e}^{-2|\alpha_j|^2}]}}.
\end{eqnarray}
\end{subequations}
The pure states $|\Psi_n\rangle$, to which Eq.~(\ref{applstate}) reduces for
$r$ $\!=$ $\!(-1)^n$, just represent the superpositions
$|0,\ldots,0\rangle\pm|1,\ldots,1\rangle$ considered in \cite{multiPlenio}.
Modeling the decoherence of the state Eq.~(\ref{applstate}) by the master
equation
\begin{equation}
\label{me}
  L\frac{\mathrm{d}\hat{\varrho}}{\mathrm{d}x}
  =\sum_{j=1}^N\eta_j(2\hat{a}_j\hat{\varrho}\hat{a}_j^\dagger
  -\hat{n}_j\hat{\varrho}-\hat{\varrho}\hat{n}_j),
\end{equation}
where $x$ may be, e.g., the propagation distance, we see that the purity
parameter $r$ and the individual complex amplitudes $\alpha_j$ decrease
according to
\begin{subequations}
\begin{eqnarray}
  r(x)&=&r(0)\mathrm{e}^{2\left(|\bm{\alpha}|^2-|\bm{\alpha}_0|^2\right)}
  \stackrel{x\ll L}{\approx}
  r(0)\mathrm{e}^{-4|\bm{\alpha}_0|^2\frac{x}{L}\eta},\qquad
  \\
\label{decoherence}
  \alpha_j(x)&=&\alpha_j(0)\mathrm{e}^{-\eta_j\frac{x}{L}}
  \hspace{0.85cm}\stackrel{x\ll L}{\approx}\alpha_j(0),
\end{eqnarray}
\end{subequations}
where $\eta$ $\!=$ $\!\frac{1}{N}\sum_{j=1}^N\eta_j$ is determined by the
$\eta_j$ $\!\ge$ $\!0$ defining the (e.g., scattering) losses and
$\bm{\alpha}_0$ $\!=$ $\!\bm{\alpha}(0)$. For distances small compared to the
classical transparency length, $x$ $\!\ll$ $\!L$, the damping of the complex
amplitudes can be neglected. For given complex amplitudes, the state
Eq.~(\ref{applstate}) is then determined by its purity parameter $r$.

The aim of our purification process is therefore to reobtain a state with
$r$ $\!=$ $\!1$ from a number of states with $0$ $\!<$ $\!|r|$ $\!<$ $\!1$.
Our model of a lossy purification is that during the feedback between subsequent
purification steps, the signal state undergoes a decoherence according to
Eq.~(\ref{me}), which leads to a damping of the purity parameter by some factor
$\eta_\mathrm{F}$ $\!\le$ $\!1$, which we call feedback efficiency. To obtain
the resulting evolution of $r$, we insert Eq.~(\ref{applstate}) into
Eqs.~(\ref{biniteration-a}), i.e.,
$\hat{\mathcal{R}}^{(j-1)}$ $\!=$ $\!\hat{\varrho}[R^{(j-1)}]$ and
$\hat{\varrho}$ $\!=$ $\!\hat{\varrho}(r)$, which gives a state
$\hat{\varrho}[R^{(j)}]$ given by Eq.~(\ref{applstate}) with
\begin{subequations}
\label{reca-a}
\begin{eqnarray}
\label{recaa}
  R^{(j)}&=&\eta_\mathrm{F}\frac{R^{(j-1)}\pm r}{1\pm rR^{(j-1)}},
  \\
  R^{(0)}&=&r.
\end{eqnarray}
\end{subequations}
Here, a positive sign corresponds to the event
\textquoteleft{{{0}}}\textquoteright\, and
negative sign to the event \textquoteleft{{{$\neg0$}}}\textquoteright\,.
With the stationary condition
$R^{(j)}$ $\!=$ $\!R^{(j-1)}$ $\!=$ $\!R^{(\infty)}$, Eqs.~(\ref{reca-a}) give a
lower and upper bound
\begin{equation}
  R^{(\infty)}=\pm\frac{\sqrt{(1-\eta_\mathrm{F})^2+4r^2\eta_\mathrm{F}}
  -(1-\eta_\mathrm{F})}{2|r|}
\end{equation}
between $|R^{(j)}|$ is able to walk. To allow an increase of $|R^{(j)}|$
above its initial value $|r|$, the condition $|R^{(\infty)}|$ $\!\ge$ $\!|r|$
must hold, which gives
\begin{equation}
  \eta_\mathrm{F}\ge\frac{1+r^2}{2}.
\end{equation}
It follows that the required feedback efficiency increases with the initial
purity. For $\eta_\mathrm{F}$ $\!\le$ $\!1/2$, the scheme is of no use at
all. On the other hand, the ideal value $R^{(\infty)}$ $\!=$ $\!1$ can only be
approached in practice. In order to keep the bound sufficiently close to one,
$|R^{(\infty)}|$ $\!\ge$ $\!1-\varepsilon$ with some given
$\varepsilon$ $\!\ll$ $\!1$, the requirement
\begin{equation}
\label{upperborder}
  \eta_\mathrm{F}\;\,_\approx^>\;1-\frac{2\varepsilon|r|}{1+|r|}
\end{equation}
must hold. For sufficiently small $\varepsilon$, a behavior similar to the
perfect case is observed. Note that the case considered here is analogous to the
example of a related two-mode state discussed in \cite{clausen10}. For a more
detailed analysis including numerical simulations, we therefore refer to that
work.
\section{\label{sec8}
       Conclusion and outlook}
We have described a procedure allowing a purification of a class of $N$-mode
quantum states as an iterative random process. While in general, a number of
identical signal state preparations is applied, only a single preparation is
required in certain cases of mixed states. The physical implementation
suggested involves beam splitter arrays, zero and single photon detections as
well as cross-Kerr elements. The role of imperfections of the procedure is
modeled in the example of a superposition of two $N$-mode coherent states. In
this case, the demands on the device increase with the initial purity of the
input state to be enhanced. A main drawback of the scheme is its relying on
large non-linearities which are difficult to implement at present. An
alternative solely based on, e.g., beam splitter arrays and photodetections
would therefore be desirable. From a theoretical point of view, the scheme
suggests that different types of entanglement may also be distinguished by the
effort required for a purification.
\begin{acknowledgments}
This work was supported by the Deutsche Forschungsgemeinschaft.
\end{acknowledgments}

\end{document}